\documentclass[gdf]{sifrev}        
\usepackage{geometry}

\newcommand{\be}{\begin{eqnarray}}
\newcommand{\ee}{\end{eqnarray}}

\begin{document}

\title{Anti-stars in the Milky Way and primordial black holes}


\author{A.D. Dolgov}           

\institute{Novosibirsk State University and BLTP, JINR, Dubna}     


\vol{42}                                  
\issue{2}				  

\month{Month year}                        



\abstract{
Astronomical data of the several recent  years which present an  evidence in favour 
of abundant antimatter population in our Galaxy, Milky Way, are analysed.
The data include: registration of gamma-rays with energy 0.511 MeV, which surely originate 
from electron-positron annihilation at rest, very large flux of anti-helium nuclei, discovered at AMS,
and 14 stars which produce excessive gamma-rays with energies of several hundred MeV 
which may be interpreted as indication that these stars consist of antimatter.
Theoretical predictions of these phenomena, made much earlier ago are described
 }
 
\maketitle

\section{Introduction} \label{sec:intro}

The story of the antimatter search in the Galaxy started probably in 1968 from the attempts initiated by
B.P. Konstantinov to search for anti-comets in the Solar System~\cite{konstantinov-1,konstantinov-2}, 
which were strongly criticised by Ya.B. Zeldovich despite very good personal relations between them.

Later activity referred mostly to cosmological antimatter but not to antimatter in our close neighbourhood,
Antimatter effects in cosmology was probably first discussed in 1971 in ref.~\cite{stecker-1}, followed
by \cite{stecker-2,chechet}. Antimatter domains in the universe were studied in \cite{khlopov-dom}.
Reviews on the state of art with antimatter in cosmology were done in \cite{stecker-2,ad-blois}.
Antimatter in the Galaxy was considered only in the last reference~\cite{ad-blois} based on the
theoretical prediction~\cite{DS}, which was later elaborated in~\cite{DKK}. 


\section{Anti-evidence}   \label{sec:anti}

\subsection{Cosmic positrons} \label{ss-positrons}
 
 Observation of intense 0.511 line, a proof of abundant positron population in the Galaxy. In the central region of the Galaxy electron--positron annihilation proceeds {at a surprisingly high rate,} creating the 
 flux~\cite{anti-e1,anti-e2,anti-e3}: 
\be\label{flux} 
\Phi_{511 \; {\rm keV}} = {1.07 \pm 0.03 \cdot 10^{-3} }\; 
{\rm photons \; cm^{-2} \, s^{-1}} .
\ee
{The width of the line is about 3 keV. 
Emission mostly goes from the  Galactic bulge and  at much lower level from the disk,}
There are several preceding works where this phenomenon was observed, 
the references can be found  in the above quoted papers.

{Until recently the commonly accepted explanation was that  ${e^+}$~\cite{AMS-21} 
are created in the strong magnetic fields of pulsars but the recent results of AMS probably exclude this 
mechanism. The reason is the following. According to the AMS data the energy spectrum of antiprotons
and positrons are exactly the same both in the form and in the absolute magnitude. This feature implies
that the mechanisms of $\bar p$ and $e^+$ production are the same and since protons cannot be produced 
in the magnetic fields of pulsars, it means that neither positrons could be created by this mechanism. 
} 
 
\subsection{Cosmic antinuclei} \label{ss-anti-nuc}

{In 2018 AMS-02 announced possible observation of six
$\overline{He}^3$ and two $\overline{He}^4$} events~\cite{AMS-18-1,AMS-18-2}.
Recent measurements reveal much more anti-events~\cite{AMS-21}.
According to the data~\cite{AMS-21}  the ratio of fluxes ${\overline{He}/He \sim10^{-9}}$ is too high,
if one assumes that  $\overline{He}$ is produced in the process of cosmic ray collisions.
As we see below, such secondary creation of $\overline{He}^4$ is negligibly weak in comparison
with theoretical expectation. A simple possibility to explain so high value of the observed anti-flux is
to assume an existence of primordial antimatter in the Galaxy. Possibly because of that
S. Ting expressed hope to observe anti-silicon,
$\overline{Si}$. 

 It is not excluded that the flux of anti-helium is even much higher than the observed one,
 because the low energy $\overline{He}$ may escape registration in AMS.

The expected rate of the secondary production of anti-nuclei in cosmic rays was calculated in ref.~\cite{anti-nuc}.
Anti-deuterium can be created in the collisions
 ${\bar p + p}$ or  ${\bar p + He}$, 
which would produce the flux of $\bar D$
${\sim 10^{-7} /m^{2}/ s^{-1}} /$steradian/GeV/neutron,
i.e. 5 orders of magnitude below the observed flux of antiprotons.
The fluxes of   
${^3\bar He}$ and ${^4\bar He}$, which could be created in cosmic ray collisions are respectively 4 and 8
orders of magnitude smaller than the flux of anti-D.
 
{After AMS announcement of observations of  anti-$He^4$ there appeared theoretical attempts to create 
anti-$He^4$ through dark matter annihilation, which does not look natural. }
A recent review on anti-nuclei in cosmic rays can be found in~\cite{anti-nuc-rev}.

\subsection{Antistars in the Galaxy} \label{ss-anti-stars}

A possible  striking discovery of antistar population in the Milky Way was announced recently~\cite{anti-stars}.
Quoting the authors:
 "We identify in the catalog 14 antistar candidates not associated with any objects belonging 
to established gamma-ray source classes and with a spectrum compatible with 
baryon-antibaryon annihilation.''

\begin{figure}
\includegraphics[scale=0.45]{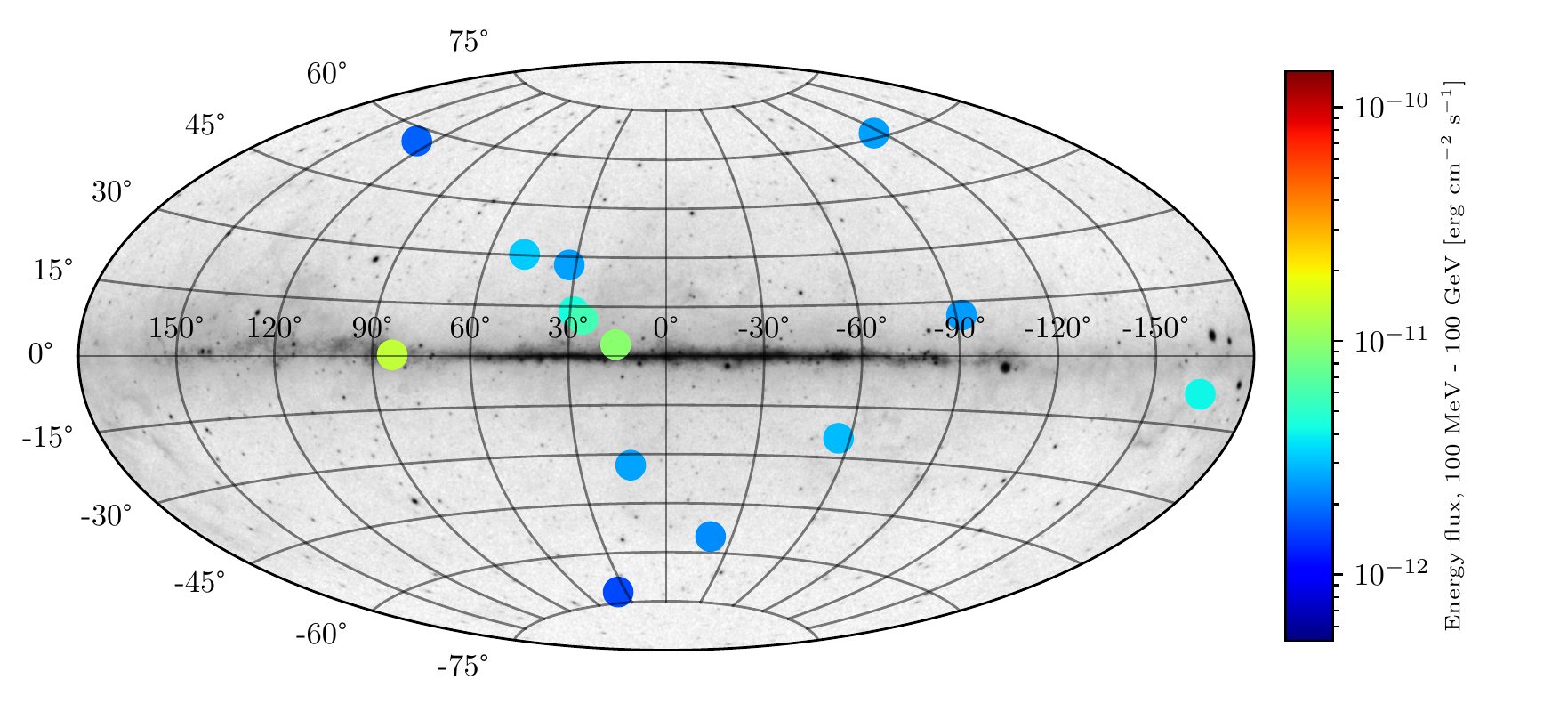}
\caption{\label{fig:sources} Positions and energy flux in the 100 MeV - 100 GeV range of antistar candidates selected in 4FGL-DR2. Galactic coordinates. The background image shows the Fermi 5-year all-sky photon counts above 1 GeV }
\end{figure}  

Of course additional confirmation of this result is necessary. A possible new way to identify antistats was
suggested in ref.~\cite{X-rays}. It was noticed there that prior to idirect $p\bar p$ contact and annihilation
the formation of atomic-like excited states consisting from proton-antiproton, proton-antinucleous
(or antiproton-nucleous), and at last nucleous-antinucleous could be formed,
which are similar to positronium.

In astrophysically plausible cases formation of such quasi-atoms  is possible in the process
of the interaction of neutral atmospheres or winds from antistars with ionised interstellar gas, 
These atoms rapidly cascade down to low levels prior to 
annihilation giving rise to a series of narrow lines which can be associated with the hadronic 
annihilation gamma-ray emission. The most significant are L (3p-2p) 1.73 keV line (yield more 
than 90\%) from ${p \bar p}$ atoms, and M (4-3) 4.86 keV (yield $\sim 60$\%) and L (3-2) 11.13 
keV (yield about 25\%) lines from $He^4 \bar p$ atoms. These lines can be probed in dedicated 
observations by forthcoming sensitive X-ray spectroscopic missions XRISM and Athena and in 
wide-field X-ray surveys like SRG/eROSITA all-sky survey.

The search of such objects could be 	facilitated by some unusual properties of antistars, e.g. by high
velocities or unusual chemical content, see below. According to the model of antistar-in-the-Galaxy 
prediction~\cite{DS,DKK} such peculiarities are typical for antistars and can be good
signatures  to search for them.			


\subsection{Observational bounds on galactic antimatter} \label{ss-obs-bounds}

Possible evidence of admixture of antimatter in the Galaxy i surely beyond the usual expectations
Normally one may expect galaxies consisting purely either of matter or antimatter. Cosmologically large
domains of matter and antimatter may be formed if C and CP symmetries are broken spontaneously. As
a result the world would be symmeric with respect to matter and antimatter.

From the data on the cosmic gamma rays one can conclude that the nearest {anti-galaxy} could not be closer 
than {at $\sim$10 Mpc}~\cite{Steigman-1}. Otherwise annihilation with protons from the common intergalactic cloud
consisting of matter would be too intensive, violating existing observational limits.

{The fraction of antimatter in Bullet Cluster} should be below ${ < 3\times 10^{-6}}$~\cite{Steigman-2}
based on  the upper bounds
to annihilation gamma-rays from galaxy clusters.
Some limits on the fraction of antimatter at large scales can be  obtained  from the CMB data
which excludes {\it large} isocurvature fluctuations at ${ d> 10}$ Mpc, and from Big Bang Nucleosynthesis
which does not allow large chemistry fluctuations at $d> 1 $ Mpc.

According to the ref.~\cite{Ballmoos-2014}
the analysis of the intensity of gamma rays created by the 
Bondi accretion of interstellar gas to the surface of an antistar would create the luminosity 
\be 
 L_\gamma \sim 3\cdot 10^{35} (M/M_{\odot})^2 v_6^{-3} .
\label{L-gamma}
\ee
It  allows to put  a limit on the relative density of antistars in the Solar neighbourhood:
{${N_{\bar *} / N_{*} <  4\cdot 10^{-5} }$} inside 150 pc from the Sun. 

{The presented above bounds are valid, if antimatter makes the 
same type objects as the observed matter.}
{For example, compact stellar-like objects consisting of antimatter may be abundant
in the Galaxy but still escape observations.
The bounds on the density of galactic compact antistars are rather loose, because the annihilation proceeds 
only on the surface of antistars  which are the objects with short mean free path of protons, as it is
analysed in the papers~\cite{antistars-1,antistars-2,antistars-3}.
 
\section{Antistar prediction  \label{s-predict} }

Based on the conventional approach no antimatter object is expected to be in the Galaxy.
However, it was predicted in 1993 and elaborated in 2009 that noticeable amount of antimatter, even 
antistars might be present in the Galaxy and in its halo in non-negligible amount~\cite{DS,DKK}.

The mechanism the papers \cite{DS,DKK} was originally dedicated to the formation of primordial 
black holes (PBH). It 
predicts a large population of massive and 
supermassive PBH with log-normal mass spectrum in perfect agreement 
with observations. This is the only known to us mass spectrum of PBH which  is tested by
observational data.

The proposed mechanism~\cite{DS,DKK}
also allows to solve multiple problems related to the observed black holes  in the
universe in all mass ranges, in particular, it explains
the origin of supermassive BHs and black holes
with intermediate masses, from
${M \sim 10^2  M_\odot}$.  up to ${10^5 M_\odot}$,  mysterious otherwise.

{As a by-product of this mechanism compact stellar type objects, which are not massive enough to form 
BHs, made of matter and antimatter are predicted.}  

\subsection{Predicted mass spectrum of PBH \label{ss:mass-spctrm}}

{The log-normal mass spectrum is determned by three constant parameters and
has the following very simple form:} 
\be
\frac{dN}{dM} = \mu^2 \exp{[-\gamma \ln^2 (M/M_0)]. }
\label{log-norm}
\ee
The parameters $\mu$ and $\gamma$ are determined by an unknown high energy physics but
$M_0$ should be equal to the cosmological horizon mass at the QCD phase transition, so
it is predicted to be ${M_0 \approx (10-20) M_\odot}$~\cite{AD-KP-M0}. This value perfectly fits the data.
In partociular, the log-normal form of the mass spectrum of PBHs} is confirmed
by the chirp mass distribution of the LIGO/Virgo events~\cite{chirp}. 
The available data on the chirp mass distribution of the black holes in the coalescing binaries in 
O1-O3 LIGO/Virgo/ runs are
analysed and compared with theoretical expectations based on the hypothesis that these black holes are 
primordial with log-normal mass spectrum. 
{The inferred best-fit mass spectrum parameters, 
{$M_0=17 M_\odot$ and ${\gamma=0.9}$,} fall 
within the theoretically expected range and shows excellent agreement with observations.} 
{On the opposite, binary black hole 
models based on massive binary star evolution require additional adjustments to reproduce the observed chirp mass distribution, see figs.~\ref{fig:chirp},\ref{fig:astro}

\begin{figure}[htbp]
\begin{center}
\includegraphics[scale=0.15]{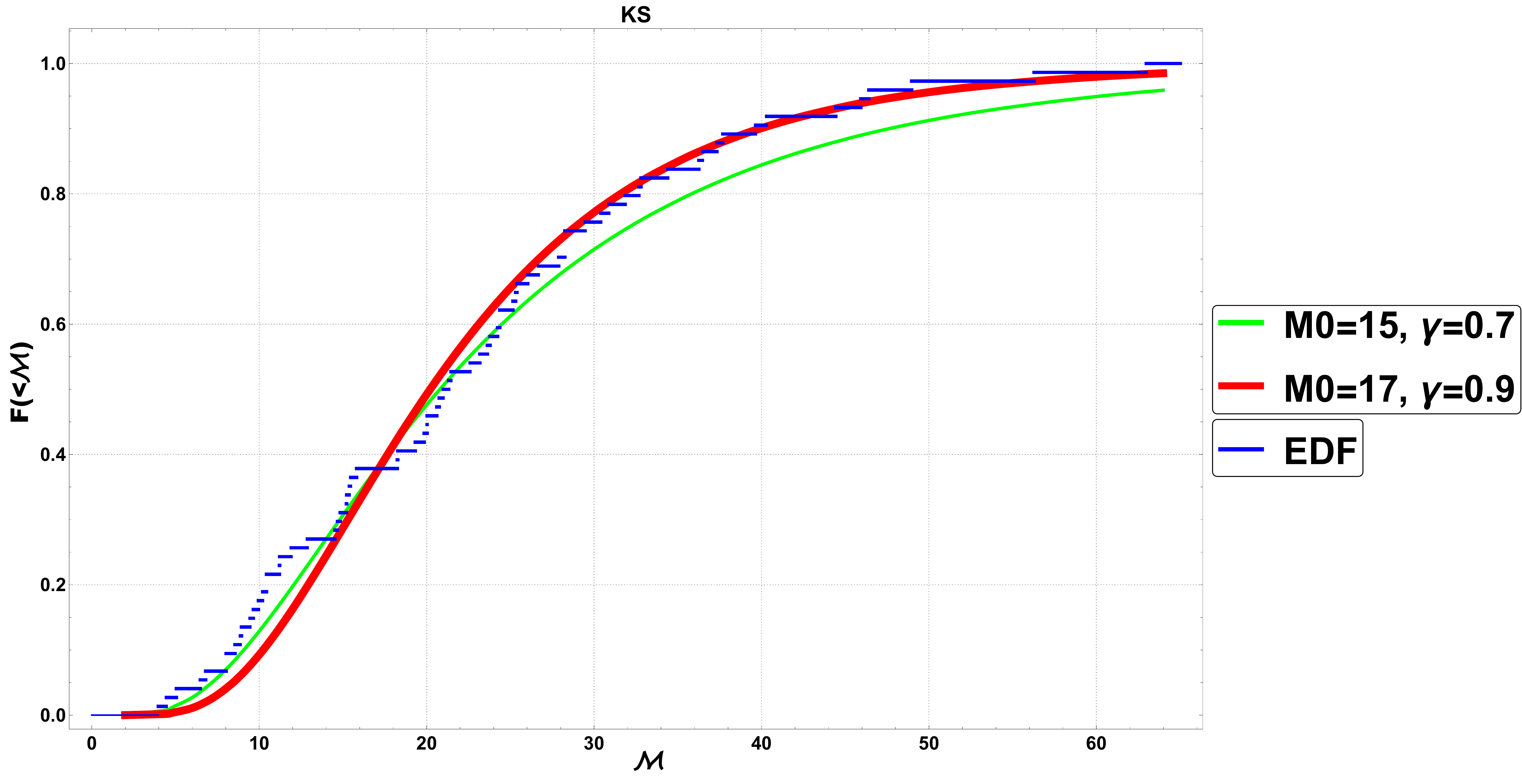}
\caption
{Model distribution $F_{PBH}(< M)$ with parameters  $M_0$ and ${\gamma}$ for two best 
Kolmogorov-Smirnov tests.  EDF= empirical distribution function.}
\label{fig:chirp}
\end{center}
\end{figure}

\begin{figure}[htbp]
\begin{center}
\includegraphics[scale=0.15]{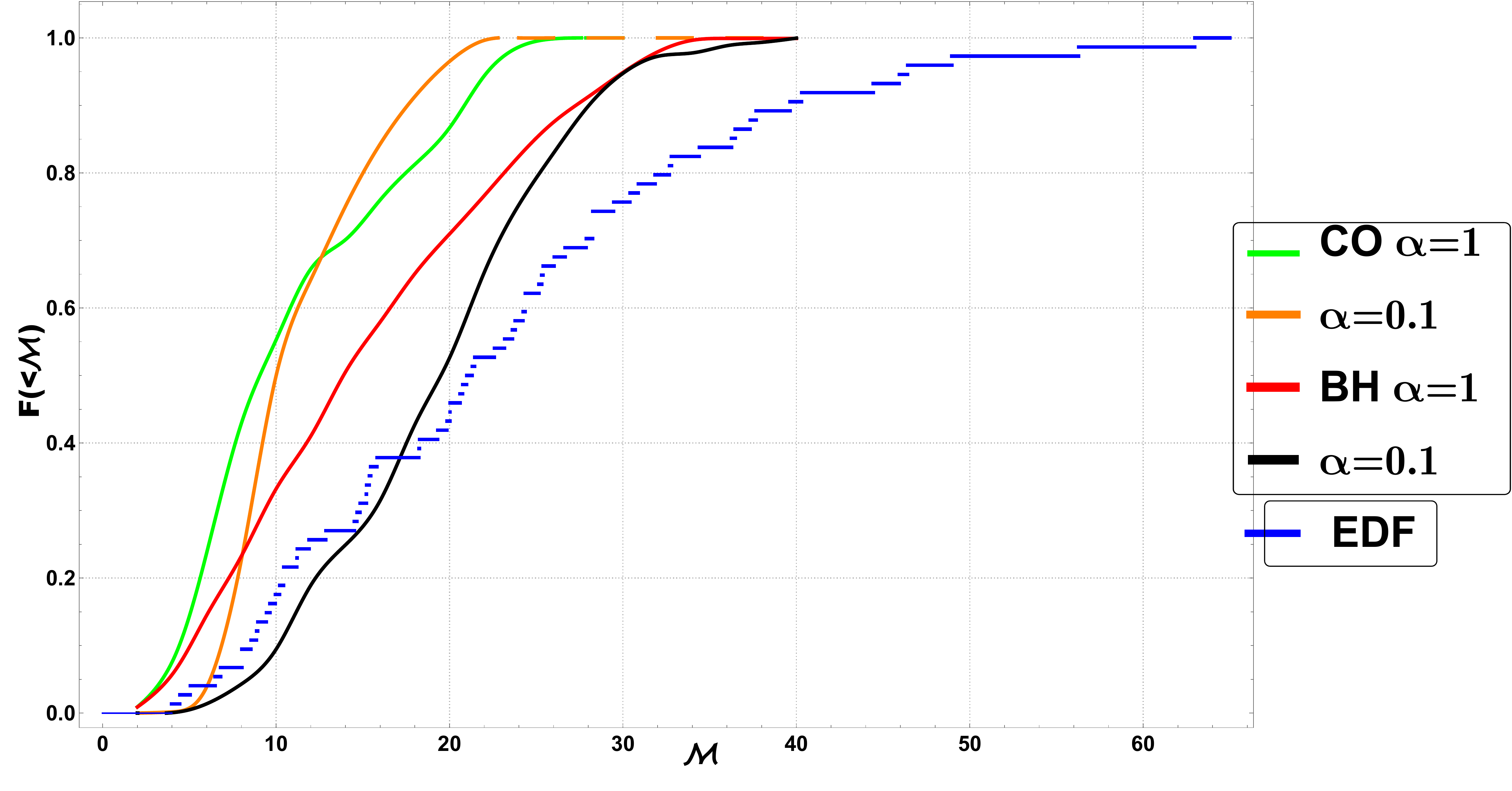}
\caption{Cumulative distributions $F(< M)$ for some  astrophysical models of binary BH coalescences. }
\label{fig:astro}
\end{center}
\end{figure}

Thus,  PBHs with log-normal mass spectrum perfectly agree with the data obtained by LIGO/Virgo/Kagra, while
astrophysical BHs seem to be disfavored. This agreement presents 
a very strong support in favour of the theoretical model~\cite{DS,DKK}.

\subsection{How reliable the galactic antimatter prediction \label{ss-reliability } }

To summarise: why can we trust the prediction the population of antimatter in the Milky Way?
In short the answer is because the it is a by-product of the mechanism of PBH formation which
pretty well resolves multiple problems with the properties of observed BH in the early ($Z\sim 10$)
and the present day universe, for a review see~\cite{AD-UFN}. 

In particular, PBHs formed according to our scenario explain the peculiar features of the sources
of GWs observed by LIGO/Virgo/Kagra~\cite{BDPP}.	

The existence of supermassive black holes (SMBH) discovered  in all large and some small galaxies and even in
almost empty environment is explained. Conventional models of SMBH formation demand time which is 
not enough by about two orders of magnitude.

The  present day universe is full of SMBH with masses in the range ${M= (10^6 - 10^{10}) M_\odot }$
as well as the intermediate mass black holes (IMBH), with $M=(10^2 - 10^5)  M_\odot $
in unexpectedly high amount. 
Moreover the SMBH are abundant in the 
the early, ${z=5-10}$, universe.
It is problematic
to explain their formation through the canonical astophysical mechanisms but much more natural
is to assume that they are primordial.

{ Since the predicted features of PBH in all mass ranges 
 well agree with the data,  one may expect that the underlying mechanism of their creation
 indeed operated in the early universe.}
 This perfectly working model of PBH creation also predicts existence of primordial antimatter in our vicinity,
 so it is quite natural to expect that antimatter is indeed abundant in the Galaxy.
 
 Theory predicts that the primordial
 stellar type compact objects have quite unusual chemical content enriched with metals, It is
confirmed in particular, by observations of extremely  old stars, even of a star which
 formally is older than the universe.  The stars looks too old because their initial chemistry is 
 enriched by heavy elements, which in the standard approach demand very long time for their 
 synthesis. Also very high velocity stars can be present in the Galaxy and they are indeed observed.

\section{Mechanism of anti-creation \label{s-anticreation}}

The suggestion of PBH (and antistar) creation proposed in references~\cite{DS,DKK}
is based on the SUSY motivated Affleck-Dine (AD)~\cite{AD} scenario of baryogenesis.
It is assumed that there exists a scalar field
$\chi$ with non-zero baryonic number, as suggested by supersymmetry.  
The potential of $\chi$ has, as a rule, the so called flat directions along which the potential
does not change, remaining the same as in the origin at $\chi = 0$.
All these properties are generically inherent to high energy supersymmetric models.

The field $\chi$ could reach very large magnitude moving along such flat directions
either due to rising quantum fluctuations
of massless fields at de Sitter (inflationary) stage, as in the original version of the AD baryogenesis model
or, as in the present paper, because of negative effective mass squared of $\chi$ (higgs-like effect).

Our new input to the model is is an introduction of the interaction of $\chi$ with the inflaton field, such that
the mass $m_{eff}^2$ stays negative for some time.
Due to this new interaction, the gate to the flat direction were closed during almost all inflationary epoch,
except for a rather short period. When the gates are closed $\chi$  remained small, sitting near 
the origin. When the gates were open $\chi$ might reach large values and when the gates close 
again, $\chi$ returned
to the origin carrying a large baryonic number but only inside the bubbles of cosmologically small but possibly
astrophysically large size.

Thus the universe would look like the inverted Swiss cheese. 
In almost all space baryon number density is small except for relatively small bubbles
with large $B$. Since the particles which carry baryonic number, i.e, quarks were massless,
the density contrast remained practically unnoticeable. The contrast became huge after the QCD phase 
transition when massless quarks turns into heavy protons and neutrons. Then these High-B Bubbles (HBB) 
turns into PBH or, if not massive enough,
into compact (anti) stars. Depending upon the direction of the $\chi$ phase rotation in the complex $\chi$ plane such
primordial stars or antistars. could be created.

When $\chi$ is large, its evolution is governed by the 
quartic part of the potential, at smaller $\chi$ the quadratic part 
dominates and if the flat directions of quartic and quadratic parts are different 
$\chi$ starts to rotate in the complex $\chi$-plane either clock-wise or anticlock-wise. As we see in what 
follows it leads either to creation of High-B-Bubble, HBB, or High-antiB-Bubbles, H${\bar B}$B.

For a toy model we assume that the quartic part of the potential has the form:
\be
U_\lambda(\chi) = \lambda |\chi|^4 \left( 1- \cos 4\theta \right),
\label{U-chi}
\ee
while the mass term can be written as ${ U_m=m^2 \chi^2 + m^{*\,2}\chi^{*\,2}}$, or in a more convenient form:
\be
U_m( \chi ) = m^2 |\chi|^2 \left[  1-\cos (2\theta+2\alpha)
\right],
\label{u-of-m}
\ee
where ${ \chi = |\chi| \exp (i\theta)}$ and ${ m=|m|e^\alpha}$.
{If ${\alpha \neq 0}$, C and CP would be  broken.} 

In GUT SUSY baryonic number is naturally non-conserved. In our toy model this non-conservation is
achieved due non-invariance of ${U(\chi)}$ with respect to phase rotation, $\chi \rightarrow e^{i \sigma} \chi$
with a constant phase $\sigma$.

{In the conventional version of the AD-baryogenesis the field $\chi$ after inflation
was away from the origin  and, when 
inflation was over, it started to evolve down to the equilibrium point, ${\chi =0}$,
according to the equation of  Newtonian mechanics with the Hubble friction term:}
\be
\ddot \chi +3H\dot \chi +U' (\chi) = 0.
\label{ddot-chi}
\ee
Baryonic number of $ \chi$, $B_\chi =\dot\theta |\chi|^2 $,
is analogous to the mechanical angular momentum of rotation in the complex $\chi$-plane.
It is quite natural that this "angular momentum" could reach a large  value and hence the baryon asymmetry,
that is the ratio of the baryonic number density to the density of the CMBR photons could be very high
$\beta = N_B/N_\gamma \sim 1$, much larger than the observed  value $\beta_{obs} \approx 10^{-9}$.
The B-conserving decays  of ${\chi}$ would transfer its
baryonic number to that of quarks. 

{Rotation could be either clockwise or anticlockwise for different bubbles with large $ \chi$ depending upon 
relative locations of the flat directions in quartic and quadratic parts of the potential 
in the complex ${\chi}$-plane. }
Large $\chi$ lives in quartic valley, but when $\chi$ drops down, it  moves to the quadratic one starting to "rotate".
So the angular momentum or, in other words $B$, is generated by possible  different 
directions of the  quartic and quadratic valleys at low ${\chi}$.
{If CP-odd phase ${\alpha}$ is non-vanishing, both baryonic and 
antibaryonic domains might be  formed}
{with possible dominance of one of them.}
Both matter and antimatter objects may exist but it is natural to expect a global dominance of one of them,
so ${ B_{tot} \neq 0}$.
 
We assume that the Affleck-Dine field $ \chi$ has the potential:
\be 
\label{U-4}
U = {g|\chi|^2 (\Phi -\Phi_1)^2}  +
\lambda_{CW} |\chi|^4 \,\ln \left( \frac{|\chi|^2 }{\sigma^2 } \right)
+\lambda (\chi^4 + h.c. ) + (m^2 \chi^2 + h.c.), 
\ee
where the first term introduced in ref.~\cite{DS,DKK}.
is the general renormalizable coupling of two scalar fields, $\chi$ and the inflaton $\Phi$.
The logarithmic term is is the Coleman-Weinberg potential~\cite{Col-Wein}, which originates from summing 
up one loop corrections to the quartic potential.

CP would be broken, if the relative phase of ${\lambda}$ and 
$m$ is non-zero, otherwise one can
``phase rotate'' $ \chi$ and come to real coefficients, eliminating thus CP-violation.

The value of $\Phi_1$ is chosen so that the $\Phi$ reached the value $\Phi_1$ in the process 
of inflation but but relatively late, so that the duration of inflation after that made about 30-40 e-foldings. 
{When the window to the flat direction is open, near ${\Phi = \Phi_1}$, }
{the field ${\chi}$ slowly diffuses to a large value,} according to quantum diffusion
equation derived by Starobinsky, 
generalised to a complex field ${\chi}$. When $\Phi$ turns large, it evolves
classically, oscillating near the local minimum of the potential.

If the window to flat direction, when ${\Phi \approx \Phi_1}$ is open only {during 
a short period,} cosmologically small but possibly astronomically large 
bubbles with high ${ \beta}$ could be
created, occupying {a small
fraction of the universe,} while the rest of the universe has normal
{${{ \beta \approx 6\cdot 10^{-10}}}$, created 
by small ${\chi}$}.         

{The mechanism of massive PBH formation quite different from all others.}
{The fundament of PBH creation is build at inflation by making large isocurvature
fluctuations at relatively small scales, with practically vanishing density perturbations.} 

{Initial isocurvature perturbations are in chemical content of massless quarks.
Density perturbations are generated rather late after the QCD phase transition.}\\[1mm]
density objects occupying a minor fraction of the universe volume.}

The main features of the scenario can be summarised as follows:
\begin{itemize}
\item 
$\rho_\chi \ll \rho_\Phi $, even inside large $\chi$ bubbles.
\item
Bubbles with large $\chi$ occupy a small fraction of the universe volume.
\item
When $\Phi < \Phi_1$ but inflation still lasts, $\chi$ is large and oscillates fast. Hence it does not
feel shallow valleys of $m^2\chi^2$. At this stage baryon asymmetry is not generated. 
\item
Inflation ends and the oscillations of $\Phi$ heats up the universe. 
\item
Ultimately the amplitude of $\chi$ drops down, the field starts to feel $m^2$-valley, and begins to 
rotate, generating large baryon asymmetry.
\item
The picture is similar to the original AD-scenario in the universe.
\item
With the chosen values of couplings and masses the density contrast between the bubbles and the rest of the world,
before the QCD phase transition, could be rather small, at the per-cent level.

\end{itemize}

\section{Summary \label{s-sum}}

Here we summarise some properties of the mechanism~\cite{DS,DKK} 
of PBH and antistar formation which are discussed above but not only them.

\begin{itemize}
\item{}
The predicted significant antimatter population of Milky Way seems to be confirmed by observations.
\item
The Log-normal mass spectrum of PBH is verified by the numerous data and very well agrees with them.
\item
PBHs formed through the mechanism~\cite{DS,DKK}  explain the peculiar features of the sources
of GWs observed by LIGO/Virgo/Kaga.
\item
The considered mechanism solves the numerous mysteries of ${z \sim 10}$ universe: supermassive black holes,, 
early created gamma-bursters and supernovae, early bright galaxies, and evolved chemistry including  a high 
level of dust.
\item
Inverted picture of the galaxy formation is proposed when a supermassive PBH is first formed and 
subsequently seeds galaxy formation.
\item
It is predicted that globular clusters (GB) and  dwarf galaxies 
are seeded by IMBHs, so there should be an intermediate mass BH
in the center with mass about 2000 $M_\odot$ in GBs and typically somewhat higher in dwarfs. An estimate
of the density of GCs is made and agrees with the data.
\item
{A large number of the recently observed intermediate mass black holes was  predicted.}
\item
{An existence of supermassive black holes observed  in all large and some small galaxies and  in
almost empty environment is explained.}
\item
{It is claimed that a large fraction of the cosmological dark matter up to 100\% can consist of PBHs.}
\item
Strange stars with unusual chemistry, enriched with metals and high velocities are predicted and observed.
Extremely old stars are predicted and  discovered.  Even, an
"older than the universe" star is found; the old age is mimicked by the unusual initial chemistry.  
\item
In refs~\cite{DS,DKK} the new ideas of invoking inflation and Affleck-Dime baryogenesis
for PBH creation were proposed, which were later repeated in a number of subsequent works.
\end{itemize}
\newpage


				          %
%


\end{document}